\documentclass[pra,aps,amsmath,amssymb,amsfonts,superscriptaddress,twocolumn,nofootinbib,longbibliography]{revtex4-1}

\usepackage{amssymb}
\usepackage{}
\usepackage{bm,mathrsfs}

\usepackage{graphicx}
\usepackage{epsfig}
\usepackage{amsmath,bbm}
\usepackage{amsfonts,amssymb}
\usepackage{times}
\usepackage{verbatim}
\usepackage[sort&compress]{natbib}

\usepackage{amsmath}
\usepackage[colorlinks,breaklinks,linkcolor=blue,anchorcolor=blue,citecolor=blue,urlcolor=blue,
dvipdfm
]{hyperref}

\newcommand{\be}{\begin{equation}}
\newcommand{\ee}{\end{equation}}
\newcommand{\bea}{\begin{eqnarray}}
\newcommand{\eea}{\end{eqnarray}}

\def\>{\rangle}
\def\<{\langle}

\def\qed{\leavevmode\unskip\penalty9999 \hbox{}\nobreak\hfill
     \quad\hbox{\leavevmode  \hbox to.77778em{%
               \hfil\vrule   \vbox to.675em%
               {\hrule width.6em\vfil\hrule}\vrule\hfil}}
     \par\vskip3pt}
     
\begin{document}

\newtheorem{theorem}{Theorem}
\newtheorem{lemma}[theorem]{Lemma}
\newtheorem{corollary}[theorem]{Corollary}
\newtheorem{proposition}[theorem]{Proposition}
\newtheorem{definition}[theorem]{Definition}
\newtheorem{example}[theorem]{Example}
\newtheorem{conjecture}[theorem]{Conjecture}
\title{Dissipative engineering a tripartite Greenberger-Horne-Zeilinger state for neutral atoms}

\author{D. X. Li}
\affiliation{Center for Quantum Sciences and School of Physics, Northeast Normal University, Changchun, 130024, China}
\affiliation{Center for Advanced Optoelectronic Functional Materials Research, and Key Laboratory for UV Light-Emitting Materials and Technology
of Ministry of Education, Northeast Normal University, Changchun 130024, China}

\author{H. W. Xiao}
\affiliation{Center for Quantum Sciences and School of Physics, Northeast Normal University, Changchun, 130024, China}
\affiliation{Center for Advanced Optoelectronic Functional Materials Research, and Key Laboratory for UV Light-Emitting Materials and Technology
of Ministry of Education, Northeast Normal University, Changchun 130024, China}

\author{C. Yang}
\email{yangc812@nenu.edu.cn}
\affiliation{Center for Quantum Sciences and School of Physics, Northeast Normal University, Changchun, 130024, China}
\affiliation{Center for Advanced Optoelectronic Functional Materials Research, and Key Laboratory for UV Light-Emitting Materials and Technology
of Ministry of Education, Northeast Normal University, Changchun 130024, China}

\author{X. Q. Shao}
\email{shaoxq644@nenu.edu.cn}
\affiliation{Center for Quantum Sciences and School of Physics, Northeast Normal University, Changchun, 130024, China}
\affiliation{Center for Advanced Optoelectronic Functional Materials Research, and Key Laboratory for UV Light-Emitting Materials and Technology
of Ministry of Education, Northeast Normal University, Changchun 130024, China}

\date{\today}

\begin{abstract}
{The multipartite Greenberger-Horne-Zeilinger (GHZ) states are indispensable elements for various quantum information processing tasks. Here we put forward two deterministic proposals to dissipatively prepare tripartite GHZ states in a neutral atom system. The first scheme can be considered as an extension of a recent work [T. M. Wintermantel, Y. Wang, G. Lochead, \textit{et al}, \href{http://dx.doi.org/10.1103/PhysRevLett.124.070503}{Phys. Rev. Lett. \textbf{124}, 070503 (2020)}]. By virtue of the polychromatic driving fields and the engineered spontaneous emission, a
multipartite GHZ state with odd numbers of atoms are generated with a high efficiency. This scheme effectively overcomes the problem of dependence on the initial state but sensitive to the decay of Rydberg state. In the second scenario, we exploit the spontaneous emission of the Rydberg states as a resource, thence a steady tripartite GHZ state with fidelity around $98\%$ can be obtained by simultaneously integrating the switching driving of unconventional Rydberg pumping and the Rydberg antiblockade effect.}
\end{abstract}

\maketitle

\section{Introduction}\label{I}
Neutral atoms excited to Rydberg states own strong, controllable Rydberg-mediated interactions that make Rydberg-atom systems become one of the most promising and versatile platforms in the fields of quantum information processing \cite{RevModPhys.82.2313}, quantum optics \cite{prapp024008ref4,prapp024008ref12}, quantum many-body physics \cite{pra052313ref44,pra032705ref14,PhysRevLett.120.063601}, and quantum metrology \cite{prapp024059ref11,pra032705ref3,prapp024059ref12,PhysRevLett.122.053601}. This exotic feature has been intensively explored and several milestones have been put forward. A prominent example is the Rydberg blockade. Benefitting from the significant suppression of the simultaneous excitation for Rydberg atoms, it serves as the backbone not only for a two qubit controlled phase gate \cite{PhysRevLett.85.2208,pra042306ref17,RevModPhys.82.2313}, but also for entanglement generation \cite{pra052313ref39,pra052313ref41,PhysRevLett.111.033606,PhysRevA.97.032701},  quantum algorithms \cite{pra052313ref43}, quantum simulators \cite{pra052313ref44}, and quantum repeaters \cite{pra052313ref45}.
On the other hands, an opposite effect, the Rydberg antiblockade\cite{PhysRevLett.98.023002,PhysRevLett.104.013001}, also sheds new light on fundamental questions about quantum logic gate \cite{PhysRevA.98.032306,PhysRevA.101.012347}, preparations of quantum entanglement  \cite{pra012328ref20,PhysRevA.96.042335,pra032336,Li:18}, and directional quantum state transfer \cite{PhysRevA.99.032348}. It is induced by combining Rydberg interactions with the two-photon detuning to realize the simultaneous excitation of two Rydberg atoms.
With the rapid development of quantum information, entanglement in bipartite systems has been well understood and quantified \cite{pra013845ref2}. More and more researchers begin to focus on unleashing the potential of multipartite entanglement in the context of measurement-based quantum computation \cite{pra062335ref9,pra062335ref13,pra062335ref19}, quantum error correction \cite{pra062335ref20,pra062335ref21}, quantum networks \cite{pra052302ref31,pra062335ref23,pra052302ref33},  and condensed matter physics \cite{pra062335ref32,pra062335ref33}. Compared with bipartite entanglement, multipartite entanglement is more powerful to manifest the nonlocality of quantum physics \cite{pra013845ref2,pra052302ref28}.

As a representative genuine multipartite entanglement, GHZ states \cite{prl260502ref6} enable a new understanding to research the local and realistic worldview further with more refined demonstrations of quantum nonlocality. Besides, they supply efficient manners for large-scale cluster state generation of measurement-based quantum computing \cite{pra052302ref37,pra052302ref42}, quantum metrology \cite{pra012337ref14,pra052302ref44,pra062335ref31}, and high-precision spectroscopy \cite{pra012337ref15,pra012337ref16}. Therefore, the preparation and measurement of GHZ states via diverse systems  have been sought for a long time and remains an attractive field of research. Nowadays,  a myriad of theoretical and experimental literatures to generate GHZ states have been proposed \cite{PhysRevLett.117.040501,PhysRevA.96.062315,pra012337ref18,oe.27.20874}.
Particularly in Ref.~\cite{PhysRevA.96.062315}, the authors presented a dissipative scheme to prepare a GHZ state of three Rydberg atoms in a cavity. Although they united quantum Zeno dynamics with Rydberg antiblockade effect to depress the harmful effect from the cavity, guaranteeing the high quality of a cavity is still a challenge in experiments, and the Rydberg atoms trapped into a cavity pronouncedly increase the
experimental difficulties.

Quite recently, integrating the Rydberg interactions and dichromatic driving fields, our group \cite{PhysRevA.98.062338} discovered another fantastic effect, unconventional Rydberg pumping (URP), which is ground-state-dependent and differs from the general Rydberg blockade or antiblockade. It will freeze the system consisting of two atoms at the same ground state and excite the system with two atoms at different ground states. The remarkable effect has exhibited the spectacular potential for various quantum information processing tasks, such as the achievement of quantum logic gate and the generation of entangled states. Furthermore, it is a meritorious pillar-stone to perform the autonomous quantum error correction for avoiding the bit-flip error of GHZ states in quantum metrology.
Additionally, analogous to the dichromatic driving fields of URP, Wintermantel \textit{et al.} \cite{PhysRevLett.124.070503} recently introduced programmable multifrequency couplings in arrays of Rydberg atoms to generalize the Rydberg blockade effect and nonunitarily  prepare GHZ states. However, the corresponding system has to be comprised of even numbers of atoms, and the optimal parameters cannot guarantee a unique steady-state solution of system. {For instance, the target GHZ state $(|0\rangle^{\otimes4}+|1\rangle^{\otimes4})/\sqrt{2}$ or $(|0\rangle^{\otimes6}-|1\rangle^{\otimes6})/\sqrt{2}$ cannot be implemented from the initial states in the basis of \{ $|1100\rangle$,~$|0110\rangle$,~$|0011\rangle$ \} or \{ $|110000\rangle$,~$|110100\rangle$,~$|110110\rangle$,~$|111100\rangle$ \}.}

Since the tripartite GHZ state is the simplest GHZ state manipulated in an experiment, and the disturbances of next-nearest neighbor Rydberg atoms can be circumvented excellently in a three-particle system, we propose two reliable schemes to dissipatively achieve the tripartite GHZ state in this paper. {Our first proposal unites the polychromatic driving fields and engineered spontaneous emissions of a short-lived level to realize the dissipative preparation of a tripartite GHZ state, which can significantly compensate for the problem of dependence on the initial state. Nevertheless, once the spontaneous emission of the Rydberg state is accessed, the population of the tripartite GHZ state will steeply descend. Thus, we design the second dissipative scheme that turns the Rydberg state decay into an important resource. The decay cooperating  with the switching driving of URP and the Rydberg antiblockade successfully generates the tripartite GHZ state with a high fidelity around $98\%$. As the target state is the unique steady state of the whole system, this scheme is also independent of the certain transport time and the tailored initial state, which is the feature of dissipative entangled-state preparations.} In what follows,
we will interpret in detail the principle of the above operations.

\section{Scheme based on polychromatic driving fields}
\subsection{Physical mechanism and effective dynamics}
\begin{figure}
\centering
\includegraphics[scale=0.24]{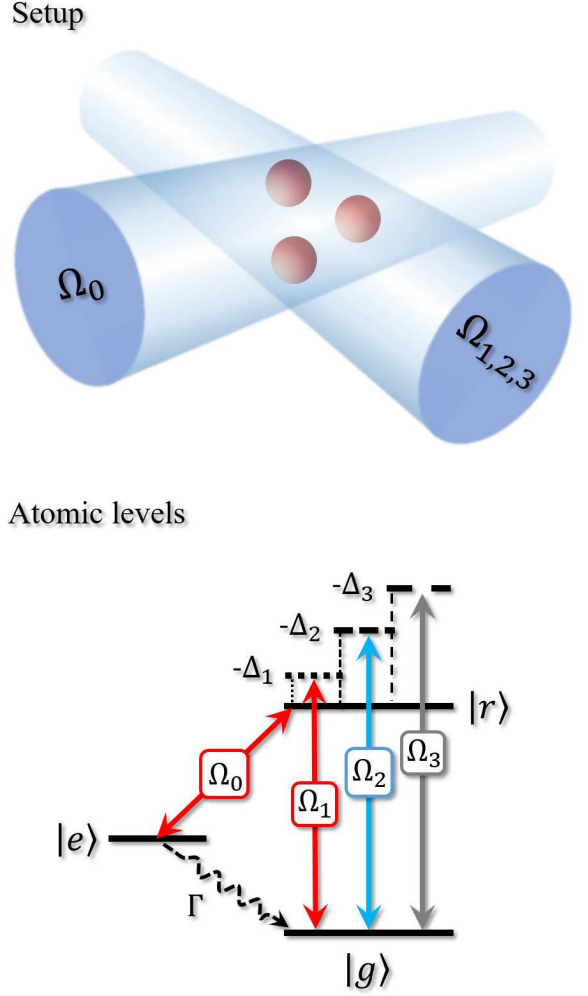}
\caption{The setup for the scheme based on the polychromatic driving fields and the engineered spontaneous emission, and the diagram of corresponding atomic energy levels. Three Rydberg atoms interact with the polychromatic driving fields $\Omega_{1,2,3}$ and a classical laser $\Omega_0$.}
\label{model1}
\end{figure}
The setup and the corresponding atomic energy levels of the scheme based on the polychromatic driving fields and the engineered spontaneous emission are illustrated in Fig.~\ref{model1}. We assume  three identical Rydberg atoms, all consisting of a ground state $|g\rangle$, a Rydberg state $|r\rangle$, and a temporary (short-lived) level $|e\rangle$, interact with the polychromatic driving fields $\Omega_{1,2,3}$ and a classical laser $\Omega_0$. While the polychromatic driving fields $\Omega_{1,2,3}$ respectively drive the transitions $|g\rangle\leftrightarrow|r\rangle$ with detunings $-\Delta_{1,2,3}$, the classical laser $\Omega_0$ resonantly couples the short-lived state $|e\rangle$ with the Rydberg state $|r\rangle$.

Supposing the three atoms decay from the short-lived state $|e\rangle$ to the ground state $|g\rangle$ with the same spontaneous emission rate $\Gamma$, the full master equation in the interaction picture can be written as
\begin{eqnarray}\label{fullm1}
\dot\rho=-i[H,\rho]+\sum_{j=1}^3L_j\rho L_j^\dag-\frac{1}{2}(L_j^\dag L_j\rho+\rho L_j^\dag L_j),
\end{eqnarray}
where
\begin{eqnarray}
H&=&\sum_{j,\alpha}\Omega_\alpha\sigma_j^{rg}e^{-i\Delta_\alpha t}+\Omega_0\sigma_j^{re}+{\rm H.c.}+\sum_{k>j}U_{jk}\sigma_j^{rr}\sigma_k^{rr},\nonumber
\end{eqnarray}
and $L_j=\sqrt{\Gamma}\sigma_j^{ge}$.
Here $|x\rangle_j\langle y|$ is parametrized as $\sigma_j^{xy}$. And $U_{jk}$ bridges the Rydberg interaction, caused by the dipole-dipole potential or the long-range van der Waals interaction, between the $j$- and $k$-th Rydberg atoms, which can obey the relation $U_{12}=U_{23}=U_{13}=U$ through the appropriate adjustments of the interatomic distance and the atomic principal quantum numbers \cite{PhysRevA.77.032723,PhysRevA.91.043802}. Since the lifetime of the temporary state $|e\rangle$ is short, we consider the decay rate $\Gamma$ is  much greater than the coupling strength $\Omega_0$, \textit{i.e.,} $\Gamma\gg\Omega_0$. And in the limiting condition of $U\gg\Omega_{0,1,2,3}$, we can reformulate the Hamiltonian in a rotating frame with respect to $U_0=\exp\{ -it \sum_{k>j}U_{jk}\sigma_j^{rr}\sigma_k^{rr}\}$,
\begin{eqnarray}\label{eff1}
H_{ I}&=&H_r+H_e,
\end{eqnarray}
with
\begin{eqnarray}
H_r&=&\sum_{j,m,n}2^{|m-n|}\Omega_{m+n+1}P_{j-1}^m\sigma_j^{rg}P_{j+1}^ne^{i(m+n-1)\Delta_1t}+{\rm H.c.},\nonumber\\
H_e&=&\sum_{j}\Omega_0P_{j-1}^0\sigma_{j}^{er}P_{j+1}^0+{\rm H.c.},\nonumber
\end{eqnarray}
where $m,n=0,1$, $P_j^0=|g\rangle_j\langle g|$, $P_j^1=|r\rangle_j\langle r|$, periodic boundary conditions of $j$ is considered, and  we have set $U=\Delta_2=(\Delta_3+\Delta_1)/2$ and $\Delta_1=2\Omega_2$ to achieve the Rydberg antiblockade effect. As for the other terms, we have neglected them as the large detuning conditions and the short lifetime of state $|e\rangle$.

The corresponding operators of atomic spontaneous emission can be simplified as $L^{(j)}=\sqrt{\Gamma}P_{j-1}^0\sigma_j^{ge}P_{j+1}^0$.
Then we can adiabatically eliminate the state $|e\rangle$ to obtain an engineered spontaneous emission.  For the sake of a clear show about the mechanism, we discard the effective Hamiltonian $H_r$, and take the Hamiltonian $\Omega_0\sigma_1^{er}P_2^0P_3^0+{\rm H.c.}$ of $H_e$ and the effective Lindblad operator $L^{(1)}$ as an example. A reduced master equation reads
\begin{eqnarray}\label{subL1}
\dot\rho_{ e}&=&-i[\Omega_0\sigma_1^{er}P_2^0P_3^0+{\rm H.c.},\rho_{ e}]\nonumber\\
&&+L^{(1)}\rho_{ e} L^{(1)\dag}-\frac{1}{2}(L^{(1)\dag} L^{(1)}\rho_{ e}+\rho_{e} L^{(1)\dag} L^{(1)}).
\end{eqnarray}
The density operator can be written in the basis of $\{|ggg\rangle,|egg\rangle,|rgg\rangle \}$ as
\begin{eqnarray}
\rho_e=\left(\begin{array}{ccc}
\rho_{gg} & \rho_{ge} &\rho_{gr}\\
\rho_{eg} & \rho_{ee} &\rho_{er}\\
\rho_{rg} & \rho_{re} &\rho_{rr}
\end{array}\right).
\end{eqnarray}
Substituting it into the Eq.~(\ref{subL1}), we can obtain a set of coupled equations for the matrix elements
\begin{eqnarray}
\dot\rho_{gg}&=&\Gamma\rho_{ee},\\
\dot\rho_{ge}&=&i\Omega_0\rho_{gr}-\frac{\Gamma}{2}\rho_{ge},\\
\dot\rho_{gr}&=&i\Omega_0\rho_{ge},\\
\dot\rho_{ee}&=&i\Omega_0(\rho_{er}-\rho_{re})-\Gamma\rho_{ee},\\
\dot\rho_{er}&=&i\Omega_0(\rho_{ee}-\rho_{rr})-\frac{\Gamma}{2}\rho_{er},\\
\dot\rho_{rr}&=&i\Omega_0(\rho_{re}-\rho_{er}).
\end{eqnarray}
In the limit of $\Gamma\gg\Omega_0$, it is reasonable to presume $\dot\rho_{ge}=\dot\rho_{ee}=\dot\rho_{er}=0$. We can solve that
$\rho_{ge}=2i\Omega_0\rho_{gr}/\Gamma$, $\rho_{er}=-2i\Omega_0\rho_{rr}/(\Gamma^2+4\Omega_0^2)$, and $\rho_{ee}=4\Omega_0^2\rho_{rr}/(\Gamma^2+4\Omega_0^2)$. Then the coupled equations of the matrix elements can be rewritten as
\begin{eqnarray}
\dot\rho_{gg}=-\dot\rho_{gg}=\Gamma_{\rm eff}\rho_{rr},~\rho_{gr}=-\frac{\Gamma_{\rm eff}}{2}\rho_{gr},~\Gamma_{\rm eff}=\frac{4\Omega_0^2}{\Gamma}.\nonumber
\end{eqnarray}
Then Eq.~(\ref{subL1}) can be derived as
\begin{eqnarray}\label{subL2}
\dot\rho_{ e}=L^{1}_{\rm eff}\rho_{ e} L^{1\dag}_{\rm eff}-\frac{1}{2}(L^{1\dag}_{\rm eff} L^{1}_{\rm eff}\rho_{ e}+\rho_{e} L^{1\dag}_{\rm eff} L^{1}_{\rm eff}),
\end{eqnarray}
where $L^1_{\rm eff}=\sqrt{\Gamma_{\rm eff}}\sigma_1^{gr}P_2^0P_3^0$. The other terms of $H_e$ and the Lindblad operators $L^{(2,3)}$  can be simplified via the similar method. Thus, the total system can be equivalent to
\begin{eqnarray}\label{effm1}
\dot\rho=-i[H_r,\rho]+\sum_{j}L^j_{\rm eff}\rho L^{j\dag}_{\rm eff}-\frac{1}{2}(L^{j\dag}_{\rm eff} L^j_{\rm eff}\rho+\rho L^{j\dag}_{\rm eff} L^j_{\rm eff}),\nonumber
\end{eqnarray}
where $L^j_{\rm eff}=\sqrt{\Gamma_{\rm eff}}P_{j-1}^0\sigma_j^{gr}P_{j+1}^0$  is the engineered spontaneous emission.

{To further describe the principle of this scheme,} we can diagonalize the resonant terms of $H_r$ and get that
\begin{eqnarray}\label{Hr}
H_r&=&\sqrt{3}\Omega(|{\rm GHZ}_+\rangle\langle E_{1+}|+|{\rm GHZ}_-\rangle\langle E_{1-}|)e^{i\Delta_1 t}+{\rm H.c.}\nonumber\\
&&+\Omega_2(2|E_{1+}\rangle\langle E_{1+}|-2|E_{1-}\rangle\langle E_{1-}|+|E_{2+}\rangle\langle E_{2+}|\nonumber\\
&&-|E_{2-}\rangle\langle E_{2-}|+|E_{3+}\rangle\langle E_{3+}|-|E_{3-}\rangle\langle E_{3-}|),
\end{eqnarray}
where we set $\Omega_1=\Omega_3=\Omega$ for simplicity and have abbreviated $|{\rm GHZ}_\pm\rangle=(|ggg\rangle\pm|rrr\rangle)/\sqrt{2}$, $|E_{1\pm}\rangle=(|grr\rangle+|rgr\rangle+|rrg\rangle\pm|ggr\rangle\pm|grg\rangle\pm|rgg\rangle)/\sqrt{6}$, $|E_{2\pm}\rangle=(|rrg\rangle-|grr\rangle\pm|rgg\rangle\mp|ggr\rangle)/2$, and $|E_{3\pm}\rangle=(2|rgr\rangle-|grr\rangle-|rrg\rangle\mp2|grg\rangle\pm|ggr\rangle\pm|rgg\rangle)/2\sqrt{3}$.
We can find that the Hamiltonian of Eq.~(\ref{Hr}) reveals the dispersive transitions of $|{\rm GHZ}_\pm\rangle\leftrightarrow|E_{1\pm}\rangle$ with detuning $\Delta_1\mp2\Omega_2$. {(Note that for the system consists of even numbers of atoms \cite{PhysRevLett.124.070503}, there is a resonant transition between $|{\rm GHZ}_+\rangle$ or $|{\rm GHZ}_-\rangle$ and a certain dark state in the presence of $\Delta_1=0$.)} Once we assume $\Delta_1=2\Omega_2$, $\Omega_2\gg\Omega$ and rotate the above Hamiltonian with $\exp\{-2i\Omega_2t(|E_{1+}\rangle\langle E_{1+}|-|E_{1-}\rangle\langle E_{1-}|)\}$, the effective Hamiltonian based on the polychromatic driving fields can amount to
\begin{eqnarray}\label{EFFH}
H_{\rm eff}&=&\sqrt{3}\Omega|{\rm GHZ}_+\rangle\langle E_{1+}|+{\rm H.c.}+\Omega_2(|E_{2+}\rangle\langle E_{2+}|\nonumber\\
&&-|E_{2-}\rangle\langle E_{2-}|+|E_{3+}\rangle\langle E_{3+}|-|E_{3-}\rangle\langle E_{3-}|),
\end{eqnarray}
where the term of $|{\rm GHZ}_-\rangle\langle E_{1-}|+{\rm H.c.}$ have been omitted as the corresponding large detuning is $4\Omega_2$ and {only the resonant transition of $|{\rm GHZ}_+\rangle\leftrightarrow|E_{1+}\rangle$ remains}. Then the effective master equation of the whole system reads
\begin{eqnarray}\label{effm}
\dot\rho=-i[H_{\rm eff},\rho]+\sum_{j}L^j_{\rm eff}\rho L^{j\dag}_{\rm eff}-\frac{1}{2}(L^{j\dag}_{\rm eff} L^j_{\rm eff}\rho+\rho L^{j\dag}_{\rm eff} L^j_{\rm eff}),\nonumber\\
\end{eqnarray}
According to the Eq.~(\ref{effm}), the target state $|{\rm GHZ}_-\rangle$ is the unique steady-state solution of this model, \textit{i.e.}, $H_{\rm eff}|{\rm GHZ_-}\rangle=L^j_{\rm eff}|{\rm GHZ_-}\rangle=0$. Therefore, initialized at an arbitrary state, the system can be stabilized at $|{\rm GHZ}_-\rangle$.

\subsection{Numerical results}
\begin{figure}
\centering
\includegraphics[scale=0.54]{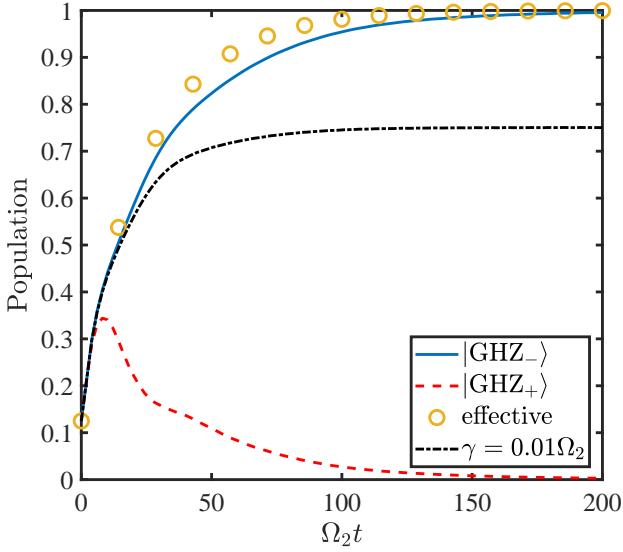}
\caption{The population of different states as functions of $\Omega_2 t$. The definition of the population for the state $|i\rangle$ is $P_i=\langle i|\rho(t)|i\rangle$. The initial states are all chosen as a mixed state $\rho_0=\sum_{l,m,n}P_{1}^lP_2^mP_3^n/8$ ($l,m,n=0,1$). The other parameters are $\Omega_0=0.77\Omega_2$, $\Omega_1=\Omega_3=0.05\Omega_2$, $\Gamma=6\Omega_2$, and $U=300\Omega_2$.}
\label{dfull}
\end{figure}

In Fig.~\ref{dfull}, we plot the dynamical evolution for the populations of the target state $|{\rm GHZ}_-\rangle$ governed by the full master equation Eq.~(\ref{fullm1}) (solid line) and the effective master equation Eq.~(\ref{effm}) (empty circles), respectively. The brilliant agreement of the two curves adequately proves the validity of the reduced system. It is significant to forecast and interpret the behaviors of the original system. Furthermore, the populations of $|{\rm GHZ}_+\rangle$ (dashed line) and $|{\rm GHZ}_-\rangle$ are respectively stable at $0.30\%$ and $99.54\%$ with the time just at $200/\Omega_2$, which reflects the feasibility and the high efficiency of the first dissipative scheme. The initial state is chosen as a mixed state $\rho_0=\sum_{l,m,n}P_{1}^lP_2^mP_3^n/8$ ($l,m,n=0,1$). It means the target state is the unique steady state of the whole system, and this is also one of the remarkable features of dissipative entangled-state preparations. {Additionally, stimulated by this principle, the present scheme can be generalized to prepare an arbitrary multipartite GHZ state with odd numbers of atoms (see the Sec.~\ref{IV} for detail).}

\begin{figure}
\centering
\includegraphics[scale=0.53]{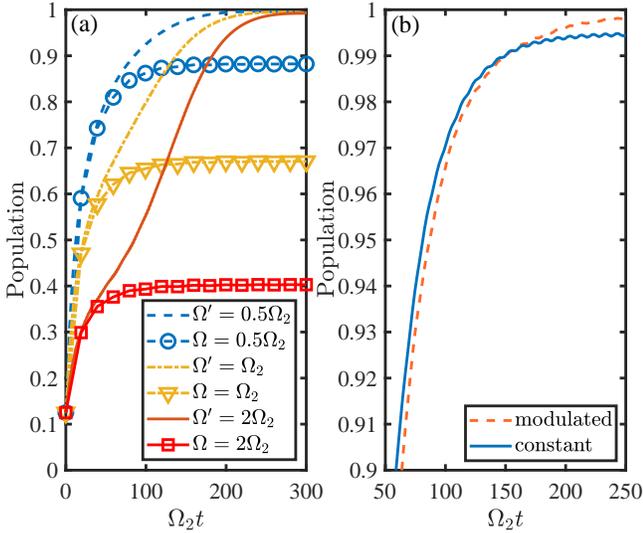}
\caption{The dynamical evolution for the population of $|{\rm GHZ}_-\rangle$ with different modulated and constant coupling strengths. The initial states and the other parameters are all the same as those of Fig.~\ref{dfull} expect for (a) $\sigma=90/\Omega_2$, $\mu=0$ and (b) $\Omega=\Omega'=0.1\Omega_2$, $\sigma=90/\Omega_2$, $\mu=110/\Omega_2$.}
\label{softduibi}
\end{figure}

In order to release the restrictive condition $\Omega_2\gg\Omega_{1,3}$, we also introduce the Gaussian pulse to improve this scheme. The key ingredient is a modulation for $\Omega_{1,3}$. In other words, the constant coupling strength $\Omega_{1,3}=\Omega$ need to be replaced into $\Omega_{1,3}(t)=\Omega'\exp[-(t-\mu)^2/(2\sigma^2)]$. Then the limiting condition of $\Omega_2\gg\Omega_{1,3}$ is no longer necessary so long as we select suitable values of $\Omega'$, $\sigma$, and $\mu$. In Fig.~\ref{softduibi}(a), we depict the populations of $|{\rm GHZ}_-\rangle$ with the polychromatic driving fields respectively applying the different modulated couplings and the corresponding constant couplings ($\Omega=\Omega'$). Owing to the Gaussian pulse, when the limiting condition $\Omega_2\gg\Omega_{1,3}$ is violated, the populations of $|{\rm GHZ}_-\rangle$ with the former still reach $99.27\%$ (dashed line), $99.35\%$ (dash-dotted line), and $99.21\%$ at $\Omega_2 t=300$.  By contrast, those with the latter markedly decrease to $88.23\%$ (empty circles), $67.07\%$ (empty triangles), and $40.29\%$ (empty squares). {On the other hand, in Fig.~\ref{softduibi}(b), when we choose $\Omega'=0.1\Omega_2,~\sigma=90/\Omega_2$, and $\mu=110/\Omega_2$, the population of $|{\rm GHZ}_-\rangle$ can be raised from $99.46\%$ (solid line) to $99.81\%$ (dashed line) at $\Omega_2t=250$. This performance manifests that the Gaussian pulse can promote the quality of the target state even though the limiting condition is not violated.}

Although the efficiency is excellent, the present scheme is sensitive to the atomic spontaneous emission of the Rydberg state $|r\rangle$, which can be described by the Lindblad operators $L_j^r=\sqrt{\gamma}\sigma_j^{gr}$ ($\gamma$ stands for the decay rate). Once we add $L_j^r$ with $\gamma$ just identical to $0.01\Omega_2$ into the Eq.~(\ref{fullm1}), the population of the target state will steeply descend from $99.54\%$ to $75.79\%$ at $\Omega_2 t=200$, which has been represented by the dash-dotted line in Fig.~\ref{dfull}. And this disadvantage is not solved in the Ref.~\cite{PhysRevLett.124.070503}, either. Consequently, we devise the second scheme based on the switching driving of URP and the Rydberg antiblockade to change the role of the Rydberg state decay into a useful resource.

\section{Scheme based on switching driving fields}
{Switching driving field is a good candidate to perfectly realize an ideal quantum process that cannot be performed by the natural evolution of systems. This technology has been used to advantage in an enormous amount of ingenious efforts,
such as the implementation of quantum logic gates \cite{PhysRevLett.82.1971,PhysRevLett.87.127901,PhysRevLett.90.217901}, the derivation and applications of the Trotter product formula $\exp\{\mathcal{L}t\}=\lim_{N\rightarrow\infty}\left(\exp\{\mathcal{L}_at/N\}\exp\{(\mathcal{L}-\mathcal{L}_a)t/N\}\right)^N$ \cite{pra012329ref33,PhysRevA.92.062114,PhysRevA.101.012329}, the preparation of entanglement with trapped ions \cite{PhysRevLett.82.1835,PhysRevA.98.042310}, and so on \cite{PRL.82.1971ref11,PhysRevA.88.044101,PhysRevB.91.064423}. In this section, we will explicate the second scheme based on the switching driving of URP in detail.}

\subsection{Physical mechanism and effective dynamics}
\begin{figure}
\centering
\includegraphics[scale=0.22]{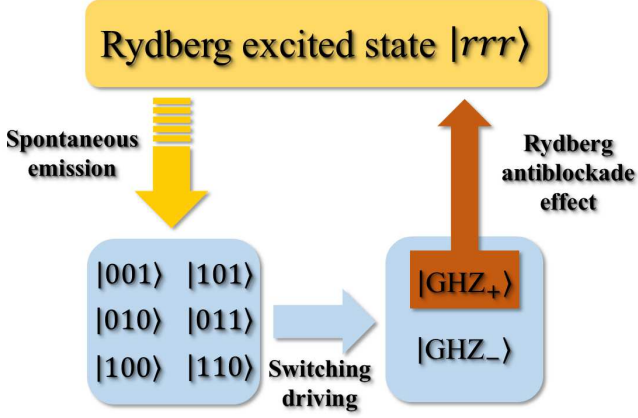}
\caption{Flow chart of the scheme based on the switching driving of URP and the Rydberg antiblockade.}
\label{flowchart}
\end{figure}

For this scheme, the system is constituted by three four-level Rydberg atoms that all encompass two ground states $|0\rangle$, $|1\rangle$ (encoded quantum bits), and two Rydberg states $|r\rangle$, $|p\rangle$. The corresponding flow chart has been elaborated in Fig.~\ref{flowchart}. It can be separated into two simultaneous processes to nonunitarily generate tripartite GHZ state $|{\rm GHZ}_-\rangle=(|000\rangle-|111\rangle)/\sqrt{2}$ with an arbitrary initial state.  One of the processes uses the switching driving of URP to transform the states with one or two atoms in state $|0\rangle$ into the subspace spanned by $|000\rangle$ and $|111\rangle$, which can be also expanded via $\{|{\rm GHZ}_+\rangle,|{\rm GHZ}_-\rangle\}$ as $|000\rangle=(|{\rm GHZ}_+\rangle+|{\rm GHZ}_-\rangle)/\sqrt{2}$ and $|111\rangle=(|{\rm GHZ}_+\rangle-|{\rm GHZ}_-\rangle)/\sqrt{2}$. In order to stabilize the system at the target state $|{\rm GHZ}_-\rangle$, the other process capitalizes on the Rydberg antiblockade effect exciting the state $|+++\rangle$ to the Rydberg excited state $|rrr\rangle$, and the stabilization of $|{\rm GHZ}_+\rangle=(|+++\rangle+|+--\rangle+|-+-\rangle+|--+\rangle)/2$ ($|\pm\rangle=(|0\rangle\pm|1\rangle)/\sqrt{2}$) can be destroyed. Subsequently, the state $|rrr\rangle$ will further decay to the ground states by the Rydberg state decay. The two simultaneous processes create a cycle among all states except $|{\rm GHZ}_-\rangle$ and lead to the system steady at $|{\rm GHZ}_-\rangle$ finally.

\begin{figure}
\centering
\includegraphics[scale=0.24]{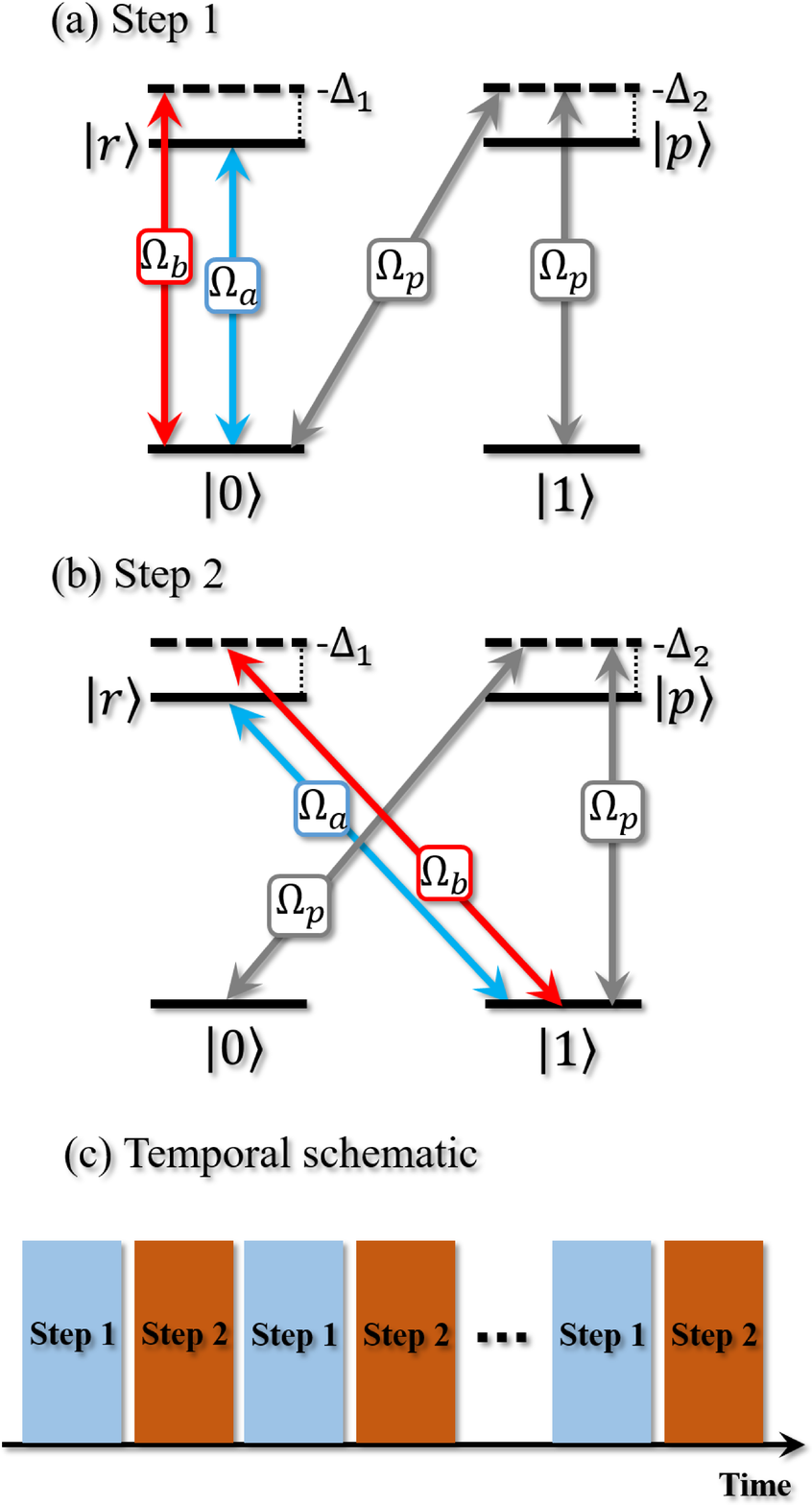}
\caption{(a) and (b) Atomic level configuration of the scheme based on the switching driving of URP and the Rydberg antiblockade, where the Step 1 and Step 2 will be carried out alternately during the whole process. The Rydberg state decays from $|r(p)\rangle$ to the ground states $|0\rangle$ and $|1\rangle$ with the same rate $\gamma_{r(p)}/2$ are not shown in the figures. (c) Temporal schematic for the alternate operations. The URP is switched frequently between the two steps, while the Rydberg antiblockade is in action all the time.}
\label{model2}
\end{figure}

In Fig.~\ref{model2}(a) and \ref{model2}(b), we flesh out the atomic levels in more detail. The process based on the switching driving of URP composes of the Step 1 and the Step 2 carried out alternately. For the Step 1, there are dichromatic driving fields with Rabi frequency $\Omega_a$ and $\Omega_b$ resonantly and dispersively (detuning $-\Delta_1$) driving the transitions $|r\rangle\leftrightarrow|0\rangle$. For the Step 2, the two lasers are switched to coupling the transitions $|r\rangle\leftrightarrow|1\rangle$ resonantly and dispersively. In the meantime, the process based on the Rydberg antiblockade effect will continuously accomplish the transitions $|p\rangle\leftrightarrow|0\rangle$ and $|p\rangle\leftrightarrow|1\rangle$ with two lasers (Rabi frequencies $\Omega_p$, detunings $-\Delta_2$) regardless of which Step in action. Moreover, we consider the Rydberg state $|r(p)\rangle$ decays to the ground states with the same rate $\gamma_{r(p)}/2$, and in what follows we set $\gamma_r=\gamma_p=\gamma$. {In Fig.~\ref{model2}(c), we also depict the temporal schematic of the alternate operations to further clarify the scheme.} In the interaction picture, the full master equation for the two steps can be written as
\begin{eqnarray}\label{jtfull1}
\dot\rho=-i[H_{S1}+H_p,\rho]+\mathcal{L}\rho,
\end{eqnarray}
and
\begin{eqnarray}\label{jtfull2}
\dot\rho=-i[H_{S2}+H_p,\rho]+\mathcal{L}\rho,
\end{eqnarray}
with
\begin{eqnarray}
H_{S1(2)}&=&\sum_{j=1}^3(\Omega_a+\Omega_be^{-i\Delta t})\sigma_j^{r0(1)}+{\rm H.c.}+\sum_{j<k}U_{rr}\sigma_j^{rr}\sigma_k^{rr},\nonumber\\
H_p&=&\sum_{j=1}^3\sqrt{2}\Omega_p\sigma_j^{p+}e^{-i\Delta t}+{\rm H.c.}+\sum_{j<k}U_{pp}\sigma_j^{pp}\sigma_k^{pp},\nonumber\\
\mathcal{L}\rho&=&\sum_{\alpha=1}^4\sum_{j=1}^3L_j^\alpha\rho L_j^{\alpha\dag}-\frac{1}{2}(L_j^{\alpha\dag} L_j^\alpha\rho+\rho L_j^{\alpha\dag} L_j^\alpha),\nonumber
\end{eqnarray}
where the Rydberg interaction of both atoms at $|r(p)\rangle$ is described by $U_{rr(pp)}$, the Rydberg interactions of two atoms occupying different Rydberg states can be ignored by means of regulating the interatomic distance and the atomic principal quantum numbers \cite{PhysRevA.77.032723,PhysRevA.91.043802}, and the Lindblad operators are $L_j^{1(2)}=\sqrt{\gamma/2}\sigma_j^{0(1)r}$ and $L_j^{3(4)}=\sqrt{\gamma/2}\sigma_j^{0(1)p}$.

\begin{figure}
\centering
\includegraphics[scale=0.19]{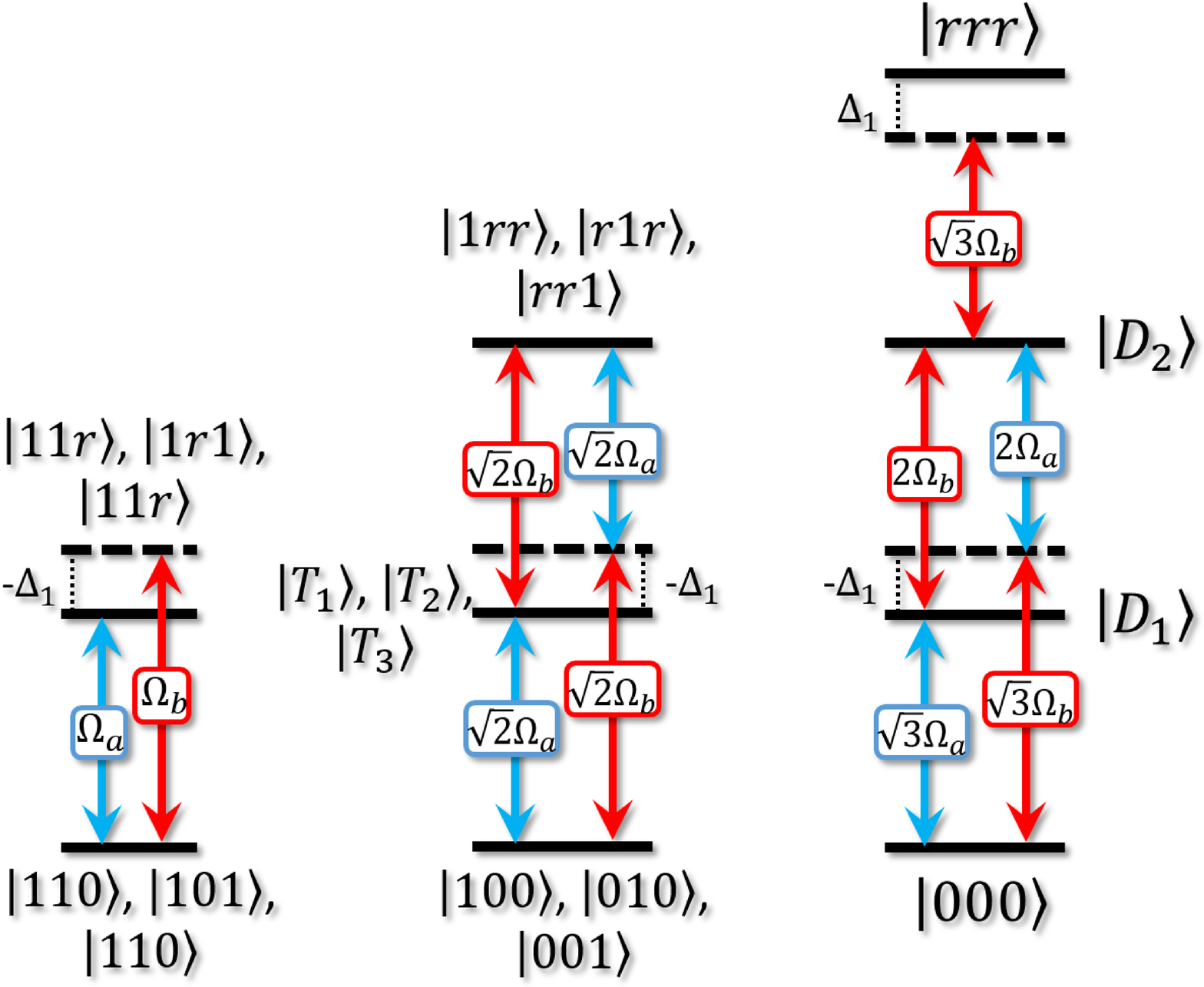}
\caption{The corresponding collective three-atom energy levels and transitions of Eq.~(\ref{HS1}).}
\label{ene}
\end{figure}
{For completeness, here we briefly reproduce some results on the URP of Ref.~\cite{PhysRevA.98.062338} that are essential to understand our computational scheme.
Referring to the conditions of the URP, we can take into account the $U_{rr}=\Delta_1$ and rotate the $H_{S1}$ with $\exp\{-it\sum_{j<k}U_{rr}\sigma_j^{rr}\sigma_k^{rr}\}$. In the large-detuning regime, the $H_{S1}$ can be divided into
\begin{eqnarray}\label{HS1}
H_{S1}&=&H_{S1}^1+H_{S1}^2+H_{S1}^3,
\end{eqnarray}
with
\begin{eqnarray}
H_{S1}^1&=&\Omega_a(|110\rangle\langle11r|+|101\rangle\langle 1r1|+|011\rangle\langle r11)+{\rm H.c.},\nonumber\\
H_{S1}^2&=&\sqrt{3}\Omega_a|000\rangle\langle D_1|+2\Omega_b|D_1\rangle\langle D_2|+\sqrt{2}\Omega_a(|100\rangle\langle T_1|\nonumber\\
&&+|010\rangle\langle T_2|+|001\rangle\langle T_3|)+\sqrt{2}\Omega_b(|1rr\rangle\langle T_1|+|r1r\rangle\nonumber\\
&&\otimes\langle T_2|+|rr1\rangle\langle T_3|)+{\rm H.c.},\nonumber\\
H_{S1}^3&=&\Big[\Omega_b(|110\rangle\langle11r|+|101\rangle\langle 1r1|+|011\rangle\langle r11)+\sqrt{2}\Omega_b\nonumber\\
&&(|100\rangle\langle T_1|+|010\rangle\langle T_2|+|001\rangle\langle T_3|)+\sqrt{2}\Omega_a(|1rr\rangle\nonumber\\
&&\otimes\langle T_1|+|r1r\rangle\langle T_2|+|rr1\rangle\langle T_3|)+\sqrt{3}\Omega_b(|000\rangle\langle D_1|\nonumber\\
&&+|rrr\rangle\langle D_2|)+2\Omega_a|D_2\rangle\langle D_1|\Big]e^{i\Delta_1t}+{\rm H.c.,}\nonumber
\end{eqnarray}
where $|D_1\rangle=(|00r\rangle+|0r0\rangle+|r00\rangle)/\sqrt{3}$, $|D_2\rangle=(|0rr\rangle+|r0r\rangle+|rr0\rangle)/\sqrt{3}$, $|T_1\rangle=(|10r\rangle+|1r0\rangle)/\sqrt{2}$, $|T_2\rangle=(|01r\rangle+|r10\rangle)/\sqrt{2}$, and $|T_3\rangle=(|0r1\rangle+|r01\rangle)/\sqrt{2}$.
To express these interactions visually, we exhibit the corresponding collective three-atom energy levels and transitions of Eq.~(\ref{HS1}) in Fig.~\ref{ene}. The ground states will be resonantly and dispersively excited to the single excited states except for the ground state $|111\rangle$ which is not evolved via $H_{S1}$. The single excited states $|T_1\rangle$, $|T_2\rangle$, $|T_3\rangle$, and $|D_1\rangle$ can be resonantly and dispersively pumped to the corresponding double excited states $|11r\rangle$, $|r1r\rangle$, $|rr1\rangle$, and $|D_2\rangle$, where the double excited state $|D_2\rangle$ will be further transferred to $|rrr\rangle$ dispersively.

In the limit of $\Delta_1\gg\Omega_b\gg\Omega_a$, $H_{S1}^3$ can approximate to the combination between the Stark-shift terms and the equivalent direct transitions from the ground states with three or two atoms at $|0\rangle$ to the corresponding double excited states. Moreover, the Stark-shift terms with the order of $\Omega_b^2/\Delta_1$ can be canceled out utilizing the other ancillary levels, while the other terms with the orders of $\Omega_a^2/\Delta_1$ and $\Omega_a\Omega_b/\Delta_1$ can be ignored as $\Omega_b\gg\Omega_a$. Consequently, $H_{S1}^3$ is useless for the scheme.}

Then we can rewrite the $H_{S1}^2$  by diagonalizing the terms of $\Omega_b$, and
\begin{eqnarray}
H_{S1}^2&=&\sqrt{\frac{3}{2}}\Omega_a|000\rangle(\langle D_+|+\langle D_-|)+\Omega_a\Big[|100\rangle(\langle T_{1+}|\nonumber\\
&&+\langle T_{1-}|)+|010\rangle(\langle T_{2+}|+\langle T_{2-}|)+|001\rangle(\langle T_{3+}|\nonumber\\
&&+\langle T_{3-}|)\Big]+{\rm H.c.}+2\Omega_b(|D_+\rangle\langle D_+|-|D_-\rangle\langle D_-|)\nonumber\\
&&+\sum_{n=1}^3\sqrt{2}\Omega_b(|T_{n+}\rangle\langle T_{n+}|-|T_{n-}\rangle\langle T_{n-}|),\nonumber
\end{eqnarray}
where $|D_{\pm}\rangle=(|D_1\rangle\pm|D_2\rangle)/\sqrt{2}$, $|T_{1\pm}\rangle=(|T_1\rangle\pm|1rr\rangle)/\sqrt{2}$, $|T_{2\pm}\rangle=(|T_2\rangle\pm|r1r\rangle)/\sqrt{2}$, and $|T_{3\pm}\rangle=(|T_3\rangle\pm|rr1\rangle)/\sqrt{2}$. According to the above equation, we can find that the effective form of $H_{S1}^2$ tends to 0 as $\Omega_b\gg\Omega_a$.
In other words, the states with three or two atoms at $|0\rangle$ cannot evolve to others by $H_{S1}^2$ since the corresponding detunings $\pm2\Omega_b$ or $\pm\sqrt{2}\Omega_b$.

To sum up, the effective Hamiltonian of $H_{S1}$ is $H_{\rm eff1}^S=H_{S1}^1$, which is the so-called URP in Ref.~\cite{PhysRevA.98.062338}. Harnessing the similar recipe, we can obtain the effective form of $H_{S2}$,
\begin{eqnarray}
H_{\rm eff2}^S=\Omega_a(|100\rangle\langle r00|+|010\rangle\langle 0r0|+|001\rangle\langle 00r|)+{\rm H.c.}.\nonumber
\end{eqnarray}

\begin{figure}
\centering
\includegraphics[scale=0.54]{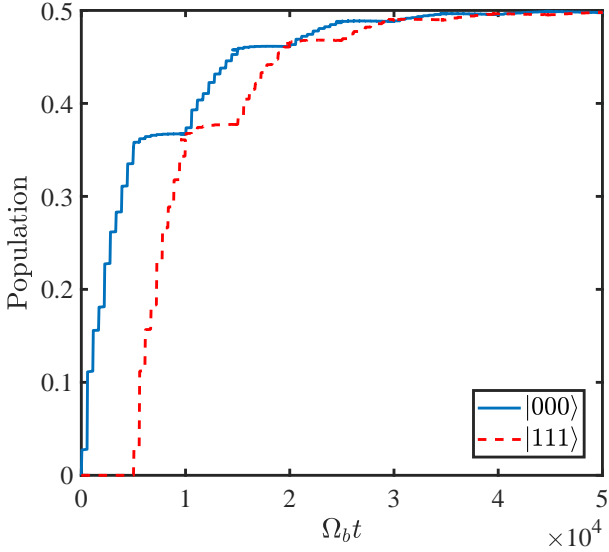}
\caption{The dynamical evolution for the populations governed by the switching driving of $H_{S1}$ and $H_{S2}$. The initial state is randomly chosen as a mixed state $\rho_0=(|100\rangle\langle100|+|010\rangle\langle010|+|001\rangle\langle001|+|011\rangle\langle011|+|101\rangle\langle101|+|110\rangle\langle110|)/6$. The other parameters are $\Omega_a=0.02\Omega_b$, $\Delta_1=300\Omega_b$, $\gamma=0.01\Omega_b$, and $N=10$ is the switching number.}
\label{firststep}
\end{figure}

In light of the switching driving of $H_{\rm eff1}^S$ and $H_{\rm eff2}^S$, it is obvious that the ground states with one or two atoms at $|0\rangle$ can be pumped into the Rydberg excited states $\{ |11r\rangle,~|1r1\rangle,~|r11\rangle,~|r00\rangle,~|0r0\rangle,~|00r\rangle \}$ that will further decay to the ground states via the spontaneous emission, and only the states $|111\rangle$ and $|000\rangle$ are steady at all times. To intuitively verify the validity of these analyses, we have plotted the dynamical evolution for the populations of $|000\rangle$ and $|111\rangle$ governed by the switching driving of $H_{S1}$ and $H_{S2}$ in Fig.~{\ref{firststep}}. After the alternate operations executed $N=10$ times, the system beginning with a mixed state is stabilized at the subspace spanned by $\{|000\rangle,|111\rangle\}$ that can be expanded via $\{|{\rm GHZ}_+\rangle,|{\rm GHZ}_-\rangle\}$. The total population trends towards unit, \textit{i.e.}, $P_{000}+P_{111}=49.79\%+49.79\%=99.58\%$ at $\Omega_b t=50000$. It faithfully designates the feasibility of the process based on the switching driving of URP.

Besides, the process taking advantage of the Rydberg antiblockade effect  makes the state $|{\rm GHZ}_+\rangle$ unstable by continuously transferring the state $|+++\rangle$ to $|rrr\rangle$. It can be indicated by the Hamiltonian $H_p$. By virtue of the basis $\{|+++\rangle,~|S_1\rangle,~|S_2\rangle,~|rrr\rangle  \}$ with $|S_1\rangle=(|++r\rangle+|+r+\rangle+|r++\rangle)/\sqrt{3}$ and $|S_2\rangle=(|+rr\rangle+|r+r\rangle+|rr+\rangle)/\sqrt{3}$, we can simplify $H_p$ as
\begin{eqnarray}
H_p&=&\sqrt{6}\Omega_p(|+++\rangle\langle S_1|+|rrr\rangle\langle S_2|)+2\sqrt{2}\Omega_p|S_1\rangle\langle S_2|\nonumber\\
&&+{\rm H.c.}-\Delta_2|S_1\rangle\langle S_1|+(U_{pp}-2\Delta_2)|S_2\rangle\langle S_2|+(3U_{pp}\nonumber\\
&&-3\Delta_2)|rrr\rangle\langle rrr|.
\end{eqnarray}
When we suppose $U_{pp}=\Delta_2\gg\Omega_p$, the Rydberg antiblockade effect is satisfied and the effective form of $H_p$ can be equal to
\begin{eqnarray}
H_{\rm eff}^p=\frac{12\sqrt{2}\Omega_p^3}{\Delta_2^2}|+++\rangle\langle rrr|+{\rm H.c.},
\end{eqnarray}
where  we have left out the order of $\mathcal{O}(\Omega_p^2/\Delta_2^2)$ and the Start-shift terms that can be canceled by ancillary levels. Due to the $H_{\rm eff}^p$, only the state $|+++\rangle$ can evolve to the state $|rrr\rangle$, which will spontaneously radiate back to the ground states.  Then the state $|{\rm GHZ}_+\rangle$ is not stable anymore. Meanwhile, combining the Rydberg state decay and the switching driving of $H_{S1}$ and $H_{S2}$, the target state $|{\rm GHZ}_-\rangle$ is turned into the unique steady state of the whole system and the second scheme is finished.

\subsection{Numerical results}
\begin{figure}
\centering
\includegraphics[scale=0.54]{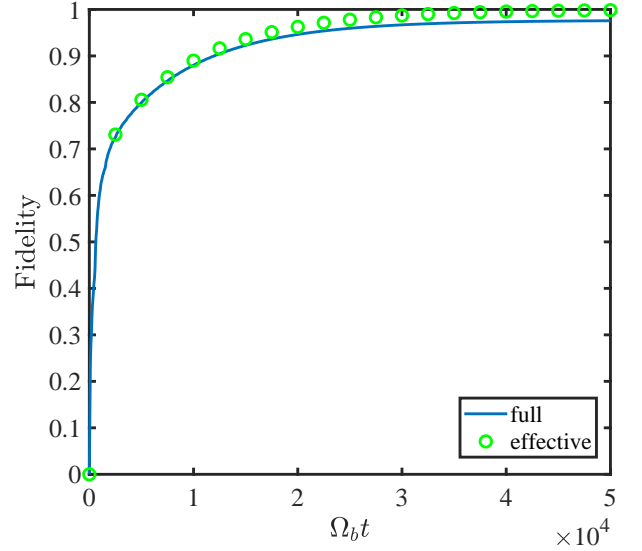}
\caption{Fidelity of $|{\rm GHZ}_-\rangle$ respectively governed by the full master equation and the effective master equation, where the definition of the fidelity  is $F={\rm Tr}\left[\rho_{\rm GHZ_-}^{1/2}\rho(t)\rho_{\rm GHZ_-}^{1/2}\right]^{1/2}$ and $\rho_{\rm GHZ_-}=|{\rm GHZ}_-\rangle\langle{\rm GHZ}_-|$. The initial state is also the mixed state $\rho_0=(|100\rangle\langle100|+|010\rangle\langle010|+|001\rangle\langle001|+|011\rangle\langle011|+|101\rangle\langle101|+|110\rangle\langle110|)/6$. The other parameters are $\Omega_a=0.02\Omega_b$, $\Omega_p=\Omega_b$, $\Delta_1=300\Omega_b$, $\Delta_2=80\Omega_b$, $\gamma=0.01\Omega_b$, and $N=64$.}
\label{full}
\end{figure}
In Fig.~\ref{full}, we characterize the fidelity of the target state $|{\rm GHZ}_-\rangle$ respectively governed by the full master equation (solid line) and the effective master equation (empty circles) in the interest of exemplifying the correctness for the above derivations, where the effective master equation can be acquired via replacing the $H_{S1(2)}+H_p$ of Eq.~(\ref{jtfull1}) (Eq.~(\ref{jtfull2})) with $H_{{\rm eff}1(2)}^S+H_{\rm eff}^p$. The empty circles is in full accord with the curve of the original system, which thoroughly certifies the rationality of the reduced system. Moreover, beginning with a mixed state $\rho_0=(|100\rangle\langle100|+|010\rangle\langle010|+|001\rangle\langle001|+|011\rangle\langle011|+|101\rangle\langle101|+|110\rangle\langle110|)/6$, the system is successfully stable at the tripartite GHZ state. It means that the present scheme is also independent of the selection of initial state. And evidently different from the previous scheme, the atomic spontaneous emission of the Rydberg states is an important tool. Thus, the fidelity can still arrive at $97.57\%$ with $t=50000/\Omega_b$ even though the rate of the Rydberg state decay reaches $0.01\Omega_b$.

\section{Discussion and conclusion}\label{IV}

\begin{figure}
\centering
\includegraphics[scale=0.54]{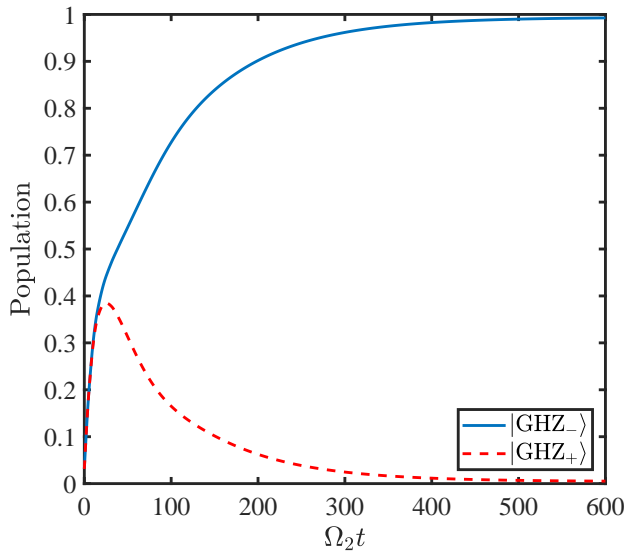}
\caption{The dynamical evolution of populations for the five-Rydberg-atom system governed by the effective master equation. The initial state is
$\rho_0=\sum_{l,m,n}P_{1}^{n1}P_2^{n2}P_3^{n3}P_4^{n4}P_5^{n5}/32$ ($n_{1,2,3,4,5}=0,1$). The corresponding parameters are $\Omega_{1,3}=0.02\Omega_2$ and $\Gamma_{\rm eff}=0.4\Omega_2$.}
\label{wu}
\end{figure}

{Here, we succinctly explain the generalization of the scheme based on polychromatic driving fields to prepare an arbitrary multipartite GHZ state with odd numbers of atoms.
For example, {we consider a five-Rydberg-atom system interacts with polychromatic driving fields $\Omega_{1,2,3}$ and a resonant laser $\Omega_0$. The corresponding atomic energy levels and transitions are the same as those in Fig.~\ref{model1}. But the next-nearest neighbor Rydberg interaction is neglected in the generalized scheme, \textit{i.e.,} the terms of $U_{jk}\sigma_j^{rr}\sigma_k^{rr}$ is replaced with $U_{j,j+1}\sigma_j^{rr}\sigma_{j+1}^{rr}$}. In the same conditions, $U_{j,j+1}=U=\Delta_2=(\Delta_3+\Delta_1)/2\gg\Omega_{0,1,2,3}$ and $\Gamma\gg\Omega_0$, a Hamiltonian similar to the $H_r$ of Eq.~(\ref{eff1}) and the engineered spontaneous emissions $L^j_{\rm eff}=\sqrt{\Gamma_{\rm eff}}P_{j-1}^0\sigma_j^{gr}P_{j+1}^0$ can be derived. To guarantee the target state $|{\rm GHZ}_-\rangle=(|0\rangle^{\otimes5}-|r\rangle^{\otimes5})/\sqrt{2}$ is the unique steady state of the system, it is the heart to set $\Delta_1=(1+\sqrt{5})\Omega_2$ which  is equal to one of the eigenvalues with respect to the resonant terms of $H_r$ of the five-Rydberg-atom system. Then we can obtain an effective Hamiltonian analogous to the Eq.~(\ref{EFFH}) in the regime of large detuning $\Omega_2\gg\Omega_{1,3}$.

In Fig.~\ref{wu}, we plot the dynamical evolution of populations for the five-Rydberg-atom system governed by the effective master equation. The feasibility of the generalized scheme is fully attested through the populations of $|{\rm GHZ}_-\rangle$ (solid line) and $|{\rm GHZ}_+\rangle$ (dashed line) respectively arriving at $99.27\%$ and $0.53\%$ with $\Omega_2t=600$. Besides, the generalized scheme needn't the certain transport time or the tailored initial state, either.}

Finally, we investigate the experimental feasibility. To date, it is an available experimental technology to arrange a group of Rydberg atoms into various geometries \cite{Schonleber_2018ref12,Ostmann_2017ref12,Ostmann_2017ref13,Schonleber_2018}. Furthermore, Ga\"{e}tan \textit{et al.} \cite{pra042306ref18} demonstrated the Rydberg interaction can be kept up to $U=2\pi\times50$ MHz between two Rydberg atoms individually trapped in optical tweezers at a distance of $4~\mu$m. And in view of this proposal, M\"{u}ller \textit{et al.} \cite{PhysRevA.89.032334} utilized two Rydberg atoms  in spatially separated dipole traps at a distance of $0.3~\mu$m to obtain a Rydberg interaction $U=2\pi\times118$ GHz and implement a controlled-$Z$ gate. Therefore, we consider the distance of the Rydberg atoms in our scheme can be varied in $[0.3,4]~\mu$m to select appropriate strengths for the interactions. In addition, the experimental realization for the couplings between the ground states and the Rydberg states actually needs a two-step transition \cite{pra052313ref41,pra042306ref18}, where the ground state and the Rydberg state will directly couple to an intermediate state dispersively. In the regime of large detuning, one can obtain the equivalent direct transition from the ground state to the Rydberg state by adiabatically eliminating the intermediate state. Accordingly, the equivalent Rabi frequency corresponding to the $\Omega_{\alpha}$ $(\alpha=1,2,3,a,b,p)$ in our scheme can be continuously tuned by the programmable Rabi frequencies and detunings of the two-step transition. Referring to the Ref.~\cite{PhysRevA.82.013405,klmoeref43,PhysRevLett.123.213606}, the decay rate of the temporary state and Rydberg states can be regarded as $\Gamma=2\pi\times5.75$ MHz (or $2\pi\times6.1$ MHz) and $\gamma=2\pi\times0.03$ MHz. In accordance with these analyses, the relationships between the relevant parameters of Fig.~\ref{dfull} and Fig.~\ref{full} are still practicable while $(\Omega_2,\Gamma)=2\pi\times (1,5.75)$ MHz and $(\Omega_b,\gamma)=2\pi\times(3,0.03)$ MHz. These reflect the experimental feasibility of the above two schemes.

To conclude, we have elaborately designed two dissipative schemes to prepare the tripartite GHZ state in a neutral atom system. In the first scheme, the GHZ states with odd numbers of atoms are successfully generated in a very short time by the organic combination between the Rydberg antiblockade effect resulting from the polychromatic driving fields and the engineered spontaneous emission induced by a temporary level. However, the first scheme is sensitive to the spontaneous emission of the Rydberg states. Therefore, in the second scheme, benefitting from the cooperation between the switching driving of unconventional Rydberg pumping caused by dichromatic driving fields and the Rydberg antiblockade effect, the spontaneous emission of the Rydberg states is characterized as a significant resource to realize the generation of the tripartite GHZ state. And the fidelity of the target state can be around $98\%$ via the state-of-the-art technology. Furthermore, because the target state is the unique steady-state solution for the whole system, the two scenarios both possess the special superiorities of dissipative entangled-state preparations, \textit{i.e.}, they never require to precisely control the transport time or exactly tailor the initial state. We believe
our schemes supply a viable prospect with regard to preparations of multipartite GHZ states.

\section*{ACKNOWLEDGMENTS}
This work is supported by National Natural Science Foundation of China (NSFC) under Grants No. 11774047.

\bibliography{URPGHZ}

\end{document}